# Ambipolar Black Phosphorus MOSFETs with Record n-Channel Transconductance

Nazila Haratipour and Steven J. Koester

*Abstract*—Ambipolar black phosphorus MOSFETs with record n-channel extrinsic transconductance are reported. The devices consist of multi-layer black phosphorus aligned to a local back-gate electrode with 10-nm-thick $HfO_2$ gate dielectric. Before passivation, devices with 0.3-μm gate length behaved as p-MOSFETs with peak extrinsic transconductance, $g_m$, of 282 μS/μm at $V_{DS}$ = −2 V. After passivation, the same devices displayed ambipolar behavior, and when tested as n-MOSFETs, had peak $g_m$ = 66 μS/μm at $V_{DS}$ = +2 V, and similar devices on the same wafer had $g_m$ as high as 80 μS/μm. These results are an important step toward realization of high-performance black phosphorus complementary logic circuits.

## I. INTRODUCTION

Black phosphorus (BP) is a promising material for high-performance transistor applications due to its high mobility and layered crystal structure which could potentially allow sub-1-nm-thick channel layers to be realized [1]. Room-temperature hole mobility as high as 1000 cm$^2$/Vs [2, 3], have been reported in few-layer BP films, and BP also has a tunable band gap which ranges from 0.3 eV in bulk to 1.0-1.5 eV for monolayers [4-6]. These properties make it an attractive alternative to other 2D materials such as graphene, which has high mobility but zero band gap [7] and transition metal dichalcogenides, which have large band gaps (1-2 eV), but low mobilities (~ 10-250 cm$^2$/Vs) [8-10]. BP also has a number of unusual properties that arise from its puckered honeycomb crystal structure including highly anisotropic electrical [4] and optical conductivity [11].

While a number of recent publications have reported high-performance BP MOSFETs [2,4,12-16], these reports have primarily involved p-channel MOSFETs (p-MOSFETs). In particular, recent papers have shown BP p-MOSFETs with transconductance of 180 μS/μm using a top-gated configuration [15], while in [16], BP p-MOSFETs using 7-nm $HfO_2$ gate dielectrics displayed transconductance as high as 250 μS/μm, drive current ~ 300 μA/μm and low contact resistance. However, despite this success, very few results on the n-channel properties of BP MOSFETs have been reported [12-14,17,18]. Since the capability to produce a complementary logic technology is critical for practical applications, development of techniques to realize high-performance BP n-MOSFETs is urgently needed.

It has previously been reported that, while unpassivated BP p-MOSFETs tend to be p-type, $Al_2O_3$ surface passivation produces a negative threshold shift [17], suggesting it could be used to produce high-performance n-MOSFETs. Similar effects of $Al_2O_3$ passivation have been observed for graphene and $MoS_2$, [19, 20]. In this letter, we report results on $Al_2O_3$-passivated BP MOSFETs fabricated using a local back-gate geometry with 10-nm $HfO_2$ gate dielectric. These devices, which have gate lengths down to 0.2 μm, show ambipolar behavior with record n-channel transconductance up to 80 μS/μm at a drain-to-source voltage, $V_{DS}$, of 2 V.

## II. DEVICE FABRICATION

The device fabrication sequence is shown in Fig. 1 and started by growing a 110-nm-thick $SiO_2$ layer on a Si substrate using dry oxidation in $O_2$. Local back gates were formed by first using electron beam lithography (EBL) to create openings in PMMA, followed by recess etching a 50-nm slot using a combination of dry and wet etching. Next, Ti (10 nm) and Pd (40 nm) were evaporated and lifted off using standard solvent processing. The gate dielectric (10 nm of $HfO_2$) was then deposited at 300 °C using atomic layer deposition (ALD). Then, black phosphorous flakes were exfoliated from the bulk

Manuscript received September 24, 2015. This work was supported by the National Science Foundation (NSF) through the University of Minnesota MRSEC under Award Number DMR-1420013. This work also utilized the University of Minnesota Nano Center and Characterization Facilities, which receive partial support from the NSF.

N. Haratipour and S. J. Koester are with the Department of Electrical and Computer Engineering, University of Minnesota, 200 Union St. SE, Minneapolis, MN 55455 (e-mail: skoester@umn.edu).

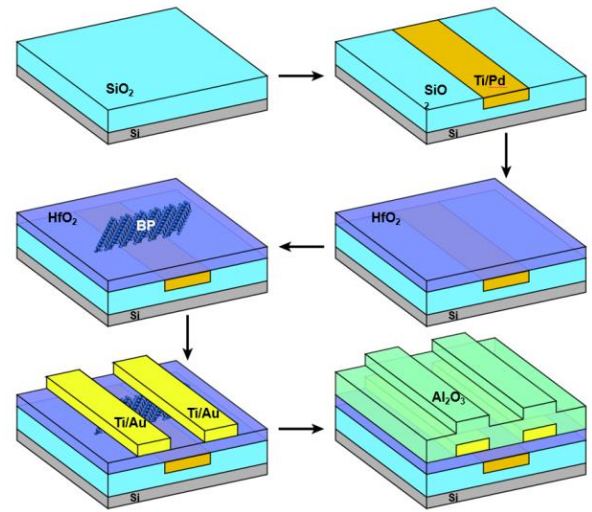

FIG. 1. Fabrication sequence for locally-backgated black phosphorus MOSFETs with $Al_2O_3$ surface passivation.

crystal and transferred onto the local back gates by using a PDMS stamp. Next, source and drain contact openings were patterned by EBL, followed by evaporation and lift off of Ti (10 nm) and Au (90 nm). Finally, the device was passivated by depositing 30 nm of $Al_2O_3$ at 200°C using ALD. The thickness of the BP flakes was determined using atomic force microscopy (AFM) on the devices after fabrication and the thicknesses were in the range of 6.4 to 13.8 nm.

## III. RESULTS

All electrical measurements were performed in vacuum at room temperature using an Agilent B1500A semiconductor parameter analyzer, and all devices were measured both before and after passivation. Before the deposition of the $Al_2O_3$ passivation layer, all of the BP MOSFETs showed strong p-type behavior. The output and transfer characteristics of one particular p-MOSFET with flake thickness of 6.4 nm and effective gate length, $L_{eff}$, of 0.3 µm before passivation are shown in Fig. 2. Here, $L_{eff}$ is defined as the spacing between the source and drain electrodes. The ON current, $I_{ON}$, is 225 µA/µm at $V_{DS} = -2$ V and $V_{GS} = -2.5$ V and the peak extrinsic transconductance, $g_m$, is 282 µS/µm for $V_{DS} = -2$ V. The latter value is actually higher than the value reported in [16], despite having a thicker gate dielectric, and to our knowledge is the highest transconductance reported for any solid-dielectric BP MOSFET.

It has been previously reported that the $Al_2O_3$ capping layer converts BP MOSFET behavior from p-type to ambipolar [17]. This shift has been attributed to the positive fixed charges in the dielectric layer [17], however further studies are needed to understand the origin of this doping mechanism. We utilized this effect and passivated the BP MOSFETs in order to analyze the electron transport in black phosphorus.

Fig. 3 shows the output and transfer characteristics of the same device as in Fig. 2 after passivation. The device shows strong ambipolar behavior so that electron transport can be probed. $I_{ON}$ reached ~35 µA/µm at $V_{DS} = 2$ V and $V_{GS} = 4$ V (Fig. 3(a)) and the peak $g_m$ is 66 µS/µm at $V_{DS} = 2$ V (Fig. 3(b)). The inset of Fig. 3(b) also shows a plot of $I_D$ vs. $V_{GS}$ for the passivated MOSFET. The ratio of the ON-current to the minimum current, $I_{ON}/I_{MIN}$, for this device is ~1.6 x $10^3$ at $V_{DS} = 0.1$ V, where the ON-current is defined as the current at $V_{GS} = V_{GS,MIN} + 2.5$ V, and $V_{GS,MIN}$ is the gate voltage where the current minimum occurs. It has been previously shown that the Ti work function lines up nearer to the BP valence band edge [13]. Therefore, at high drain bias the tunneling at the drain side can become significant, to the point where the OFF current actually exceeds $I_{ON}$. It will be important for future optimization of BP n-MOSFETs to identify lower work function metals that can produce lower Schottky barrier heights for electron injection.

Similar n-channel performance was observed in several BP

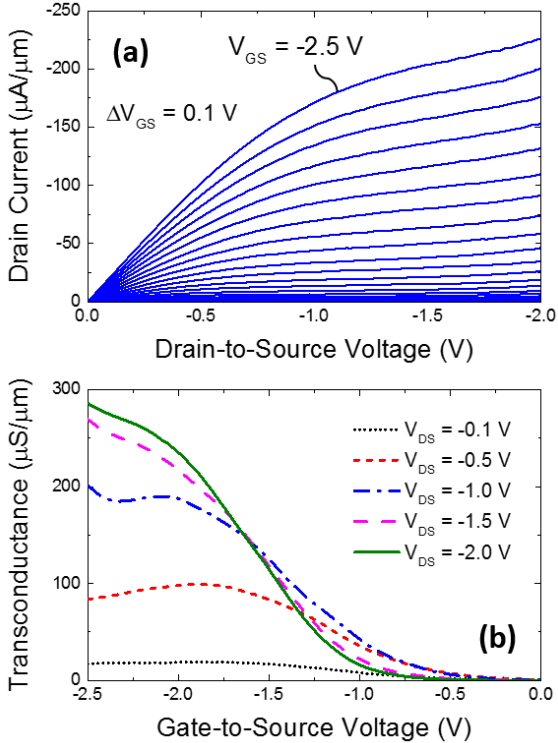

FIG. 2. (a) Drain current, $I_D$, vs. drain-to-source voltage, $V_{DS}$, at room temperature of a BP p-MOSFET with channel length of 0.3 µm and $HfO_2$ gate dielectric thickness of 10 nm. The maximum current corresponds to $V_{GS} = -2.5$ V and the gate voltage step is 0.1 V. (b) Transconductance, $g_m$, vs. gate-to-source voltage, $V_{GS}$, for the same device at $V_{DS} = -0.1, -0.5, -1.0, -1.5$ and $-2.0$ V. The maximum transconductance is 282 µS/µm at $V_{DS} = -2.0$ V.

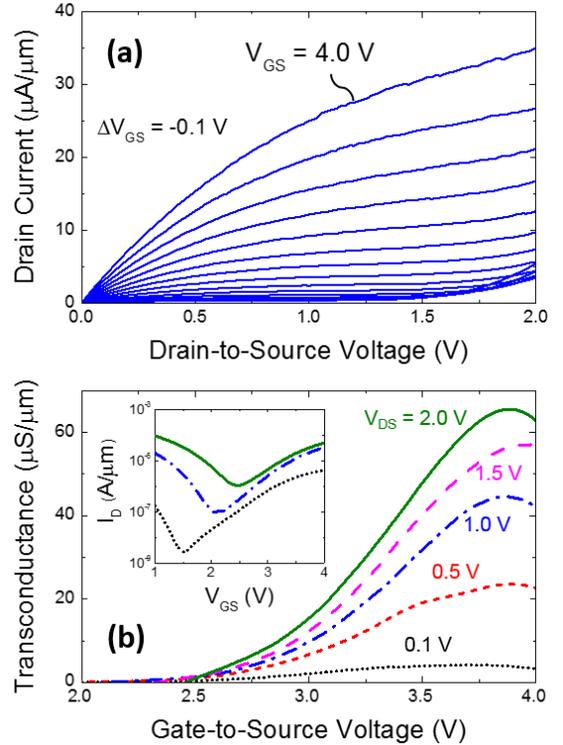

FIG 3. (a) Drain current, $I_D$, vs. drain-to-source voltage, $V_{DS}$, at room temperature for same BP MOSFET as in Fig. 2 after 30 nm $Al_2O_3$ passivation. The maximum current corresponds to $V_{GS} = +4.0$ V and the gate voltage step is 0.1 V. (b) Transconductance $g_m$, vs. gate-to-source voltage, $V_{GS}$, for the same device at $V_{DS} = 0.1, 0.5, 1.0, 1.5$ and $2.0$ V. The maximum $g_m$ is 66 µS/µm at $V_{DS} = 2.0$ V. Inset: Semilog plot of $I_D$ vs. $V_{GS}$ of the same device for $V_{DS} = 0.1, 1.0,$ and $2.0$ V.

MOSFETs, and Fig. 4(a) shows the $g_m$ vs. $V_{GS}$ characteristics for four devices fabricated on the same wafer. The best device had peak n-channel $g_m$ of 80 μS/μm which is the highest value reported to date for n-type conduction in a BP MOSFET.

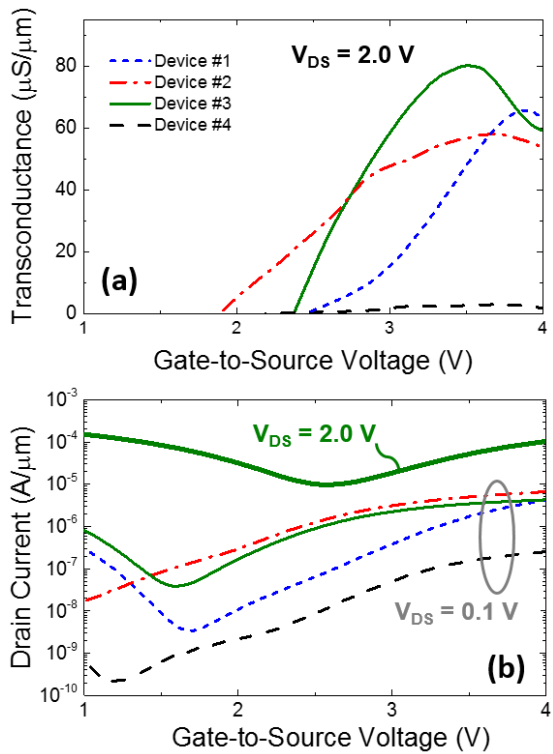

FIG. 4. (a) N-channel extrinsic transconductance, $g_m$, for different BP MOSFETs at $V_{DS}$ = 2.0 V. The maximum $g_m$ is 80 μS/μm for device #3. The effective channel length, $L_{eff}$ = 300 nm (200 nm) for devices #1, 3 (devices #2, 4). (b) $I_D$ vs. $V_{GS}$ at $V_{DS}$ = 0.1 V for the same devices as in (a). The characteristic for device #3 is shown at $V_{DS}$ = 2.0 V.

The effect of passivation on the subthreshold slope was also characterized before and after passivation. When the devices were measured as p-MOSFETs, the passivation dramatically improved the subthreshold slope from 740 mV/dec to 300 mV/dec (averaged over one decade of current for all four devices in Fig. 4). However, the subthreshold slope was degraded to 570 mV/dec for the same passivated devices when measured as n-MOSFETs. We believe the initial improvement after passivation is due to the high temperature bake-out provided by the ALD system that desorbs water from beneath the BP. However, the degraded subthreshold slope for passivated n-MOSFETs compared to p-MOSFETs could indicate that the $Al_2O_3$ passivation produces interface traps near the conduction-band side of the band gap. This hypothesis is supported by the prominent "kink" in the subthreshold behavior for three of the devices, as shown in Fig. 4(b).

Finally, a summary of the performance characteristics for the devices is compiled in Table 1. First of all, we note that the $I_{ON}/I_{MIN}$ ratio (defined at $|V_{DS}|$ = 0.1 V) scales with BP thickness, as expected based upon the thickness-dependent band gap of BP. In addition, one of the devices had substantially lower $g_m$ than the others, for both when measured as a p- and n-MOSFET. While the precise origin of this behavior is not entirely understood, we note that the crystal direction of the BP was not controlled in these experiments and that large spread in the $g_m$ values could be due to the random rotational orientation of the devices. To explore this behavior further, the effective field effect mobility, $\mu_{FE}$, was extracted using

$$\mu_{FE} = g_m L_{eff} / W C_{ox} V_{DS}, \qquad (1)$$

where $g_m$ is the peak transconductance at $|V_{DS}|$ = 0.1 V, $L_{eff}$ is the effective gate length, $W$ is the channel width, and $C_{ox}$ is the estimated gate capacitance per unit area for 10 nm of $HfO_2$ with relative permittivity of 16.6. The highest hole mobility for device #1 is roughly consistent with our prior results on $HfO_2$-gated BP p-MOSFETs [16]. In addition, the large mobility variations between devices and between electrons and holes are roughly consistent with expectations based upon crystal asymmetries [11]. Finally, we note that the contact resistance was not extracted in our mobility calculation, which can lead to significant degradation of the effective mobility, particularly given the short gate length of our devices [5].

TABLE I
SUMMARY OF DEVICE RESULTS BEFORE AND AFTER PASSIVATION

| Device index | $t_{BP}$ (nm) | $L_{eff}$ (μm) | Before passivation | | After passivation | | |
|---|---|---|---|---|---|---|---|
| | | | $\mu_{H\text{-}eff}$ (cm²/Vs) | $g_{m\text{-}PFET}$ (μS/μm) | $\mu_{N\text{-}eff}$ (cm²/Vs) | $g_{m\text{-}NFET}$ (μS/μm) | $I_{ON}/I_{MIN}$ @$V_{DS}$=0.1 V |
| 1 | 6.4 | 0.3 | 38 | 282 | 7.2 | 66 | 1.6 x 10³ |
| 2 | 10.9 | 0.2 | 11 | 179 | 5.6 | 58 | 2.9 x 10² |
| 3 | 13.8 | 0.3 | 20 | 231 | 6.0 | 80 | 1.3 x 10² |
| 4 | 9.3 | 0.2 | 5.0 | 48.7 | 0.26 | 3.0 | 1.1 x 10³ |

IV. CONCLUSION

In summary, we have demonstrated locally back-gated black phosphorus MOSFETs with 10 nm $HfO_2$ gate dielectric and $Al_2O_3$ capping layer. These devices produce record n-channel extrinsic transconductance as high as 80 μS/μm. Further performance enhancements could be obtained by improving the contact and dielectric quality. These results are encouraging for using BP MOSFET for CMOS logic circuit applications.